\documentstyle[psfig]{mn}
\renewcommand{\baselinestretch}{1.0}
\newif\ifAMStwofonts

\ifoldfss
  \ifCUPmtlplainloaded \else
    \NewTextAlphabet{textbfit} {cmbxti10} {}
    \NewTextAlphabet{textbfss} {cmssbx10} {}
    \NewMathAlphabet{mathbfit} {cmbxti10} {} 
    \NewMathAlphabet{mathbfss} {cmssbx10} {} 
  \fi
  \ifAMStwofonts
    \ifCUPmtlplainloaded \else
      \NewSymbolFont{upmath} {eurm10}
      \NewSymbolFont{AMSa} {msam10}
      \NewMathSymbol{\upi}     {0}{upmath}{19}
      \NewMathSymbol{\umu}     {0}{upmath}{16}
      \NewMathSymbol{\upartial}{0}{upmath}{40}
      \NewMathSymbol{\leqslant}{3}{AMSa}{36}
      \NewMathSymbol{\geqslant}{3}{AMSa}{3E}

       \let\le=\leqslant
       
    \fi
  \fi
\fi 

\ifnfssone
  \newmathalphabet{\mathit}
  \addtoversion{normal}{\mathit}{cmr}{m}{it}
  \addtoversion{bold}{\mathit}{cmr}{bx}{it}
  \newmathalphabet{\mathbfit} 
  \addtoversion{normal}{\mathbfit}{cmr}{bx}{it}
  \addtoversion{bold}{\mathbfit}{cmr}{bx}{it}
  \newmathalphabet{\mathbfss} 
  \addtoversion{normal}{\mathbfss}{cmss}{bx}{n}
  \addtoversion{bold}{\mathbfss}{cmss}{bx}{n}
  \ifAMStwofonts
    \ifCUPmtlplainloaded \else
      \UseAMStwoboldmath
      \makeatletter
      \new@mathgroup\upmath@group
      \define@mathgroup\mv@normal\upmath@group{eur}{m}{n}
      \define@mathgroup\mv@bold\upmath@group{eur}{b}{n}
      \edef\UPM{\hexnumber\upmath@group}
      \new@mathgroup\amsa@group
      \define@mathgroup\mv@normal\amsa@group{msa}{m}{n}
      \define@mathgroup\mv@bold\amsa@group{msa}{m}{n}
      \edef\AMSa{\hexnumber\amsa@group}
      \makeatother
      \mathchardef\upi="0\UPM19
      \mathchardef\umu="0\UPM16
      \mathchardef\upartial="0\UPM40
      \mathchardef\leqslant="3\AMSa36
      \mathchardef\geqslant="3\AMSa3E

       \let\le=\leqslant

    \fi
  \fi
\fi 

\ifnfsstwo
  \DeclareMathAlphabet{\mathbfit}{OT1}{cmr}{bx}{it}
  \SetMathAlphabet\mathbfit{bold}{OT1}{cmr}{bx}{it}
  \DeclareMathAlphabet{\mathbfss}{OT1}{cmss}{bx}{n}
  \SetMathAlphabet\mathbfss{bold}{OT1}{cmss}{bx}{n}
  \ifAMStwofonts
    \ifCUPmtlplainloaded \else
      \DeclareSymbolFont{UPM}{U}{eur}{m}{n}
      \SetSymbolFont{UPM}{bold}{U}{eur}{b}{n}
      \DeclareSymbolFont{AMSa}{U}{msa}{m}{n}
      \DeclareMathSymbol{\upi}{0}{UPM}{"19}
      \DeclareMathSymbol{\umu}{0}{UPM}{"16}
      \DeclareMathSymbol{\upartial}{0}{UPM}{"40}
      \DeclareMathSymbol{\leqslant}{3}{AMSa}{"36}
      \DeclareMathSymbol{\geqslant}{3}{AMSa}{"3E}

       \let\le=\leqslant

    \fi
  \fi
\fi 

\ifCUPmtlplainloaded \else
  \ifAMStwofonts \else 
    \def\upi{\pi}
    \def\umu{\mu}
    \def\upartial{\partial}
  \fi
\fi

\title{Gas Rich Galaxies and the HI Mass Function}
\author[J.I. Davies et al.]
       { J.I. Davies$^{*}$, W.J.G. de Blok$^{+}$, R.M. Smith$^{*}$, A. Kambas$^{*}$,  S. Sabatini$^{*}$, \and S. M. Linder$^{*}$ and 
S. A. Salehi-Reyhani$^{*}$
\\ $^{*}$ Cardiff University, Department of Physics and Astronomy, Queen's Buildings,
P.O.Box 913, Cardiff CF24 3YB, UK  \\
$^{+}$ATNF, PO Box 76, Epping, NSW 1710, Australia}
\date{Received in original form}

\pagerange{\pageref{firstpage}--\pageref{lastpage}}
\pubyear{2001}

\begin{document}

\maketitle

\label{firstpage}

\begin{abstract}
We have developed an automated cross-correlation technique to detect 21cm emission in sample spectra obtained from the HI Parkes All Sky Survey. The initial sample selection was the nearest spectra to 2435 low surface brightness galaxies in the catalogue of Morshidi-Esslinger et al. (1999). The galaxies were originally selected to have properties similar to Fornax cluster dE galaxies. As dE galaxies are generally gas poor it is not surprising that there were only 26 secure detections. All of the detected galaxies have very high values of $(M_{H}/L_{B})_{\odot}$. Thus the HI selection of faint optical sources leads to the detection of predominately gas rich galaxies. The gas rich galaxies tend to reside on the outskirts of the large scale structure delineated by optically selected galaxies, but they do appear to be associated with it. These objects appear to have similar relative dark matter content to optically selected galaxies. The HI column densities are lower than the 'critical density' necessary for sustainable star formation and they appear, relatively, rather isolated from companion galaxies. These two factors may explain their high relative gas content. We have considered the HI mass function by looking at the distribution of velocities of HI detections in random spectra on the sky. The inferred HI mass function is steep though confirmation of this results awaits a detailed study of the noise characteristics of the HI survey.

\end{abstract}

\section[]{Introduction}
It is believed that galaxies evolve via star formation from initially a gas dominated to finally a stellar (and stellar remnants) dominated state. Although the average star formation rate of the Universe possibly had a peak value somewhere between $z=1-2$ \cite{mad96,lil96}, individual galaxies can have very different star formation histories. Star formation seems to have either started at different times and/or has proceeded at different rates at different places in the Universe. 
Elliptical galaxies appear to have formed very early and converted their gas into stars very quickly whilst, galaxies like the the giant Low Surface Brightness (LSB) galaxy Malin 1 \cite{bot87,imp89} still have huge reservoirs of gas and seem to be forming stars at a (constant ?) slow rate. The globally averaged star formation rate of galaxies has been determined using rather high surface brightness galaxies, measuring either the ultra-violet and/or the far-infrared luminosity density \cite{bla99}. What is not clear from these measurements is whether there is a significant population of galaxies, similar to Malin 1, that have continued to form stars at very much lower rates over longer periods of time. These galaxies would be very difficult to detect in the ultra-violet, because of their LSB, and in the far-infrared, because of their low gas phase metalicity (hence low dust content) and because of the large average distance between the stars the dust would be very cold. Given their hypothesised large relative gas mass the most fruitfull region of the spectrum to detect them would seem to be 21cm

Malin 1 is different to other spiral galaxies in a number of ways. For example its $(M_{H}/L_{B})_{\odot}$ of 5 \cite{bot87,imp89} is much higher than the 0.1 for a 'typical' spiral galaxy and very much more than 0.01 for a 'typical' elliptical. While there does seem to be a systematic increase in $(M_{H}/L_{B})_{\odot}$ from $\approx 0.05$ for early type spirals to $\approx 1$ for very late type irregulars \cite{kna90}, galaxies with values as high as Malin 1 are quite extraordinary. A large value of $(M_{H}/L)_{\odot}$ indicates either a galaxy that is very young or one that has been forming stars at a very low rate, for its mass, compared to other galaxies. If we can find larger numbers of galaxies with these properties then we will gain a much better understanding of the factors that govern galaxy and star formation rates.

The reasons why galaxies form stars at different rates is not totally clear, but there are two very likely prime factors. These are the initial conditions (the gas density at formation) and the frequency of galactic interactions. High density and/or a large number of encounters are both thought to promote star formation. Thus elliptical galaxies formed at a place of high  initial over density and subsequently had many interactions and mergers with smaller galaxies, while Malin 1 probably formed at a spatially large, but small overdensity in a very isolated environment. By studying relatively isolated gas rich galaxies we have the opportunity of studying galaxies that have not had rapid star formation induced by interactions, undergone mergers and have not been tidally stripped. As long as they have not suffered significant expulsion or accretion of gas the mass function of these galaxies should reflect the initial mass function of galaxies at their time of formation. Gas rich galaxies are thus the best galaxies to compare with the initial density fluctuations \cite{pre74} assumed in recent numerical models of galaxy formation \cite{fre96,kau97}.

There are two fundamental ways of detecting atomic hydrogen in gas rich systems, either by absorption or emission. QSO absorption line studies indicate large numbers of gas rich systems (from the damped Ly$_{\alpha}$ systems to the Ly$_{\alpha}$ forest) most of which have no identifiable optical counterparts. 21cm observations have predominately concentrated on, and therefore almost always appear to be associated with, optical systems. In addition, the QSO data tends to sample the distant Universe while the 21cm observations have concentrated on rather nearby objects. QSO absorption line observations are generally sensitive to much lower column densities (down to $\approx10^{12}$ $atoms$ $cm^{-2}$) than 21cm observations (at best $\approx10^{18}$ $atoms$ $cm^{-2}$), but even where the two regimes overlap there are many objects that have no optical counterparts. Some damped Ly$_{\alpha}$ systems, for example, which were generally believed to arise in the discs of 'typical' large spiral galaxies have now, after much closer scrutiny, been found to arise from absorption in dwarf and LSB galaxies \cite{coh00,bow00}. If further observations confirm this for other damped Ly$_{\alpha}$ systems then we will have to move away from the view that the majority  of hydrogen absorption lines occur in huge gas haloes around 'typical' galaxies to one in which the gas is clumped into much smaller, previously undetected clouds. So we might speculate on the possibility of HI rich clouds like this existing nearby and thus being accessible to 21cm observation at column densities of $\approx10^{18}$ $atoms$ $cm^{-2}$ or above (for an alternate view see Rao \& Briggs, 1993). As 21cm observations become more sensitve and extensive we will be able to test this hypothesis. A start can be made using the first 21cm all sky survey \cite{bar01}.

So is there a large local population of galaxies that have converted only a small fraction of their gas into stars ? If so, what is their spatial distribution ? What is the form of their HI mass function and how does such a population relate to current numerical models of galaxy formation ?
To try and answer these questions we have used 21cm data taken from the HI Parkes All Sky Survey (HIPASS)
\footnote {The Parkes telescope is part of the Australia Telescope which is funded by the Commonwealth of Australia for operation as a National Facility managed by CSIRO}
to study the HI properties of a sample of LSB galaxies. The extracted HI spectra are those closest to the optical positions of the 2435 galaxies in the LSB galaxy sample of \cite{mor99a}. Where required we have used $H_{0}=75$ $km$ $s^{-1}$ $Mpc^{-1}$.

\section[]{The data}
The optical data are taken from the photographic survey for LSB galaxies carried out by Morshidi-Esslinger et al. (1999a and b). The survey covered approximately 2000 $sq$ $deg$ using data obtained from APM scans of UK Schmidt telescope survey plates. Galaxies were selected to be 'similar' to previously detected dE galaxies in the Fornax cluster. We use the word 'similar' because the APM automated detection routine is optimised to select rather smooth looking images like dE galaxies, rather than dI or spiral galaxies. The latter tend to have a 'lumpy' appearance which the APM classifier often splits into separate or 'noise' images. The photometric  selection criteria was a central surface brightness in the B band fainter than 22.5 B$\mu$ and an exponential scale length greater than 3 $arc$ $sec$. Full details of the optical survey data are given in \cite{mor99a}.

The optical data were originally used to study the total numbers, numbers in different environments and the clustering scales of LSB dwarf galaxies. Given that we tried to optimise the galaxy selection to dE galaxies we actually might not expect any HI detections. Previous observations of dE galaxies in clusters indicate HI masses of less than $10^{7}$ $M_{\odot}$ \cite{imp88} while our sensitivity (see below) is only below $10^{7}$ $M_{\odot}$ for velocities less than about 1500 $km$ $s^{-1}$. Thus we do not expect to be able to detect the numerically dominant dE galaxies in this sample. Rather, we are trying to detect 'interlopers', that is objects that appear similar to LSB dE galaxies yet contain larger amounts of HI. In fact Malin 1 was discovered in a similar way. It was originally thought to be a dwarf galaxy in the Virgo cluster, but was later discovered to be a giant LSB galaxy in the background \cite{bot87}. Our models and observations \cite{mor99a,mor99b}, indicated some 'background' contamination of the optical sample by more distant objects, not necessarily dE galaxies. We were hoping that some of these might be gas rich galaxies like Malin 1.

The HI data comes from the 388 $8^{o}\times8^{o}$ survey data cubes of the HiPASS 21cm southern sky survey \cite{bar01}. The angular resolution of the data is 15.5 $arcmin$ after the data have been gridded. The grid spacing in each cube is 4 $arc$ $min$ and the spectrum used is the nearest spectrum to the optical position. The channel spacing is 13.2 $km$ $s^{-1}$ and the velocity resolution is 27 $km$ $s^{-1}$ after smoothing. There are 1024 channels, but we initially considered only velocities in the range 400-12000 $km$ $s^{-1}$. The lower limit was set to avoid local hydrogen, the upper limit by the proximity of the velocity cut-off. The typical noise fluctuation in each spectrum after smoothing is 0.006 Jy beam$^{-1}$. The identification of sources in the HI data is described below.

\section[]{HI detection}
The automated detection to well defined selection criteria of images on, for example, CCD frames has become much more sophisticated over the last few years (see for example \cite{ber96}). Numerous computer packages exist to automatically select galaxies to well defined selection criteria (isophotal size and magnitude for example). This does not appear to be the case for HI detections yet the two problems are very similar. For example Kilborn (2001) discusses an automated galaxy finder for use on HiPASS data cubes, but then resorts to selection by eye. In none of the papers on blind HI surveys, we have come across, do we find an objective selection criteria for HI sources \cite{zwa97,sch98}. These papers supply information about the observing setup and the data reduction, but say little about the detection of objects from the spectra obtained. In the main objects appear to be identified by eye and there is no explanation of the selection criteria except to say (incorrectly) that there is some lower mass limit at each distance.  

We have previously been involved with techniques for the detection of LSB galaxies in imaging data \cite{phi93,dav94,kam00}. A number of years ago it had become clear that LSB galaxies were very much underrepresented in optically (by eye) selected samples taken from imaging data. The lesson learnt was that only when a full analysis of the selection process had been carried out could you then define the sorts of galaxies you would and would not be able to detect (\cite{dis76,dis83,dav90}. Carrying out deeper observations with better understood selection criteria has led to the detection of numerous LSB galaxies.

An optical image of a galaxy is detected against a systematically varying background level with the addition of random noise fluctuations. Detection of the HI signal is very similar - the varying background is the base-line ripple and in addition there are random noise fluctuations. For optical image detection there is no well defined magnitude or size limit - sample selection is always a combination of magnitude and size. For example one can always think of a galaxy that is bright enough to be part of a magnitude limited sample, but fails to get in because it is to large (its surface brightness is less than or close to the survey isophotal limit). In a similar way large velocity width galaxies with low central intensities will be missed or asigned to base line ripples even though they contain sufficient hydrogen, in total, to be detected in a 'mass limited' survey.  In this section we describe how we have applied some of our previous techniques of surface photometry to the detection of HI sources.
 
\begin{figure}
\centerline{\psfig{figure=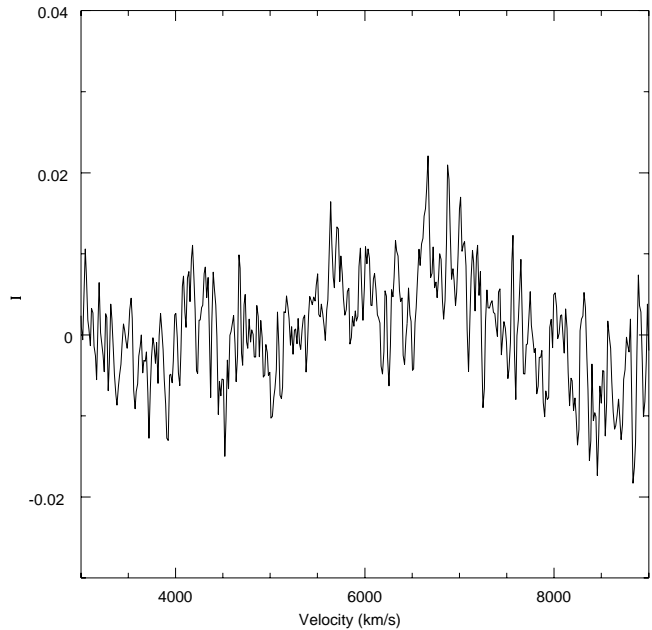,width=9cm}}
\caption
{A 3$\sigma$ fluctuation in a HI spectra at $\approx6700$ $km$ $s^{-1}$}.
\label{fig:3sig}
\end{figure}

\begin{figure}
\centerline{\psfig{figure=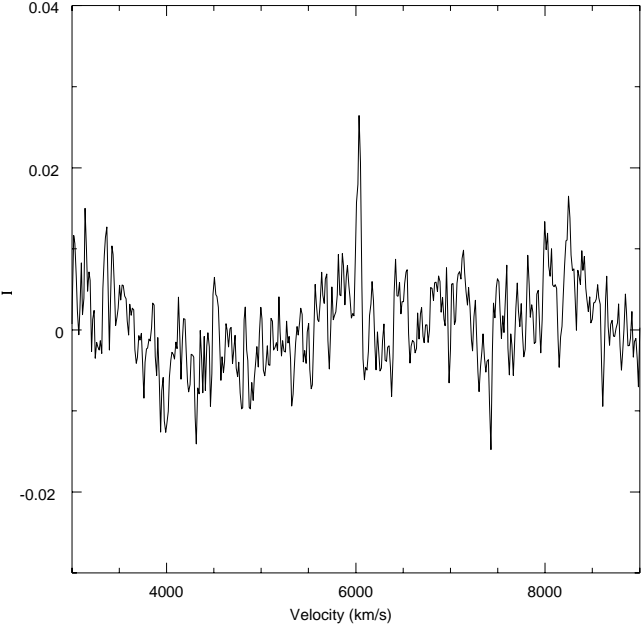,width=9cm}}
\caption
{A 4$\sigma$ detection with a velocity width that is just resolved ($\Delta v \approx 38$ $km$ $s^{-1}$)}.
\label{fig:4sig}
\end{figure}

Having 2435 spectra to inspect was another strong motivation for employing an automated technique.  As mentioned above there are two important factors that influence our ability to detect 'HI objects' in HI spectra. The first is random noise the second is baseline fluctuations. The signal is the integral over the line width, so large signals can arise from large peak values and/or large velocity widths. The problem with identifying large velocity widths without large peak values is that they can look the same as baseline fluctuations (see also section 6 and figure~\ref{fig:noise}). The problem with the random noise is that the expectation (Gaussian) is one single channel 3$\sigma$ ($\sigma$ is the standard deviation of the data values) fluctuation in the 1000+ channels. Thus 3$\sigma$ detections (see figure~\ref{fig:3sig}) are not reliable unless they have sufficiently large velocity widths, but even then, if the velocity width is too large, they can resemble baseline fluctuations. By 'hiding' simulated galaxies in real spectra it became clear that an initial 4$\sigma$ detection was required, because even 3$\sigma$ peak values with quite large velocity widths were not convincingly different to the noise. So the initial object identifier was simply one of peak value at the 4$\sigma$ level. At 4$\sigma$ we would expect one false detection in every 30 spectra or about 80 false detections in the sample as a whole. So the second requirement was that the initial peak value detection also had a 'resolved' velocity width (figure~\ref{fig:4sig}). That is a velocity width greater than 27 $km$ $s^{-1}$. With this criteria we would not expect any detections by chance (see also section 6 below). With these criteria the lowest signal to noise ratio for detection  is about 10, a value that would be readily accepted in imaging data. 

Our selection criteria does not lead to an integrated flux limited sample (see above), so we will use the term  'survey limits' to indicate our two minimal selection criteria. This is analogous to what would be referred to (incorrectly) as the magnitude limit for a magnitude limited imaging survey sample. There are two other points. Firstly our selection criteria will lead to the preferential selection of face-on, rather than edge-on, disc galaxies as these will have higher central intensities and narrower line profiles. Secondly 'spikes' in the data like that illustrated in figure~\ref{fig:4sig} are very similar in velocity width but, lower in amplitude, to the confirmed detection of an apparently isolated HI cloud by Kilborn et al. (2000).  

Given the noise in the data it is difficult to measure the velocity width and flux integral accurately. To minimise this problem we have cross-correlated the data with templates and used the best fitting template to derive the central velocity, velocity width and flux integral. 
Inspection of a small part of the data indicated that by far the majority of the sources appeared as single 'spikes' rather than 'double horned'. The exact form of the template used is not critical, but the maximum gain in signal to noise is obtained for an 'optimum filter', this is one that has the same shape as the object being measured, this is an example of matched filtering \cite{irw90}. We have choosen Gaussian templates as they appear 'similar' to the profile shapes we are trying to measure and they are relatively easy to interpret. The cross-correlation program will also measure 'double horned' profiles (but not so accurately) by fitting Gaussians around the central velocity.

Essentially we use a technique similar to one we have used before to detect and measure LSB galaxies \cite{dav94,phi93}. We have used this method to derive the best fitting exponential central surface brightness and scale length of LSB galaxy images (photometry). Here we will derive the central intensity and velocity width of the best fitting Gaussian to the HI spectrum. The cross-correlation technique for surface photometry is fully described in \cite{phi93} (PD). Below we will briefly describe our method using similar notation to PD.

The correlation coefficient of a spectrum $G_{s}$ and a model template spectrum $G_{t}$ is defined in general by the convolution

\begin{equation}
C_{st}(r)=\frac{\int G_{s}(x)G_{t}(x+r)dx}{( \int G_{s}^{2}(x) dx)^{1/2} ( \int G_{t}^{2}(x) dx)^{1/2}}
\end{equation}

where $r$ is the shift between the spectrum and template and the integrals are taken over their intersection. In practice the integrals are taken as sums over the digitised data. $C_{st}^{2}\le1$ regardless of the form of $G_{t}$ or $G_{s}$ and there is a maximum when $G_{t}=\alpha G_{s}$, where $\alpha$ is a positive constant (see PD). If the data substantially exceeds the scale size of the Gaussians used we can ignore the limits on the integrals of equation 1.  So, if the velocity profile of the template and galaxy are both Gaussians then $G_{t}(v)=I_{t}\exp{-(v/\alpha_{t})^{2}}$ and $G_{s}(v)=I_{s}\exp{-(v/\alpha_{s})^{2}}$, where $v$ and $\alpha$ are velocities and I is an intensity. Substituting into 1 and integrating gives

\begin{equation}
C_{st}=\sqrt{\frac{2\alpha_{s}\alpha_{t}}{\alpha_{s}^{2}+\alpha_{t}^{2}}}
\end{equation}

which obviously has a maximum when $\alpha_{s}=\alpha_{t}$. By convolving with different Gaussian templates over a range of velocities we can determine the best matching template and the central velocity from the maximum value of $C_{st}$. To determine the flux integral we also need to know the central intensity. To do this we use a second convolution, but with a different normalisation

\begin{equation}
A_{st}(x)=\frac{\int G_{s}(x)G_{t}(x)dx}{ \int G_{s}^{2}(x) dx}
\end{equation} 

For the Gaussian case this reduces to 

\begin{equation}
A_{st}=\frac{I_{t}}{I_{s}}\sqrt{\frac{2 \alpha_{t}^{2}}{\alpha_{s}^{2}+\alpha_{t}^{2}}}
\end{equation}

which implies $A_{st}=\frac{I_{t}}{I_{s}}$ when $\alpha_{s}=\alpha_{t}$. So once we have selected the correct template ($\alpha_{t}$), $A_{st}$ is just the ratio of the unknown intensity $I_{s}$ to the known normalization of the model, $I_{t}$.

The velocity profile will not be a perfect Gaussian so the cross-correlation will find a closely matching Gaussian model. This will be the 'best' fitting model in the sense of minimizing the weighted sum of the deviations.

In practise we will always have a noisey image. For example, if the noise per pixel is everywhere Gaussian with a fixed amplitude $\sigma$ (this assumes the signal is not large compared to the noise) then

\begin{equation}
\int G_{s}^{2}(v) dv=\int ( I_{s}\exp{-(v/\alpha_{s})^{2}} + N )^{1/2} dv 
\end{equation}

while the other terms remain unchanged. N represents a Gaussian random error term with mean zero and standard deviation $\sigma$. As the cross-terms have an expectation value of zero equation 
2 becomes

\begin{equation}
C_{st}=\sqrt{\frac{2\alpha_{s}\alpha_{t}}{( \alpha_{s}^{2}+\alpha_{t}^{2}) (1+\frac{ \sqrt{2}x_{t}\sigma^{2}}{\sqrt{\pi} \alpha_{s} I_{s}^{2}})}}
\end{equation}

where $x_{t}$ is the intersection of the spectrum with the template. The 'correction' term involves only the parameters of the noise and the spectrum and so the maximum still occurs at $\alpha_{s}=\alpha_{t}$. As one might expect the effects of the noise are minimised for large values of $\alpha_{s} I_{s}^{2}$ (large velocity widths and peak values). During the cross-correlation process we kept the intersection, $x_{t}$, the same for each template so that we could compare the correlation coefficients of different templates. 

We have used Gaussian templates with full width half maximum values from 25 to 500 $km$ $sec^{-1}$ at 12.5 $km$ $sec^{-1}$ intervals. We reject those that are not velocity resolved ($v<27$ $km$ $s^{-1}$). The largest  detected velocity width in the sample is 337.5 $km$ $s^{-1}$ some way below the largest template size. The detection process has been fully tested on a wide range of simulated and real data. Using simulated Gaussian profiles, in real data, with central intensities of 4$\sigma$ we find that we can estimate profile parameters and HI masses to about a factor of 3. As confirmation of this one of the galaxies in the final sample (see below) has a previous 21cm measurement. For F300-026 Matthewson and Ford (1996) measure an HI mass of $8\times10^{8}$ $M_{\odot}$ (using our derived distance of 11.3 $Mpc$) while the value derived from our cross-correlation program is $6\times10^{8}$ $M_{\odot}$. We have also excluded regions in each spectra that contain known sources of noise (HIPASS web page).

In summary the automated HI detection process involved firstly the identification of a 4$\sigma$ or higher value, then finding the maximum correlation coefficient with a template of velocity width (full width at half-maximum) greater than 27 $km$ $s^{-1}$, the velocity resolution of the data (in practice the width of the smallest 'resolved' template of 37.5 $km$ $s^{-1}$). This is what we define as our 'sample limits'.

After carrying out the template matching we had a list of 155 HI  detections. We then needed to carry out other checks to see how secure these detections were. The main problem is one of reliably asigning the optical and HI detections to the same object. Typically the galaxies in the optical sample have diameters of 0.3 $arc$ $min$. The HI resolution is 15.5 $arc$ $min$. To overcome this problem we have used the optical Digital Sky Survey (DSS) to inspect the area around each HI detection. We have also used the NASA/IPAC Extra galactic Database (NED) to find known objects within 10 $arc$ $min$ of the position of the HI spectra. We have removed objects from our initial list if NED has a similar redshift for another object within 10 $arc$ $min$ or if there is a more prominent galaxy within the field of view. For strong nearby signals ($v<2000$ $km$ $s^{-1}$) we searched NED for galaxies of similar redshift at up to 1 $deg$ away (by looking at spectra at random positions around nearby galaxies (such as NGC1365 and NGC1291) it is clear that they can affect spectra up to 30 $arc$ $min$ away from their optical centre). As an indication that some of our optical and HI detections come from the same object we can compare the measured HI velocities with previously determined (optical) velocities obtained from NED, these are available for 10 galaxies in our sample (see table 1). In all cases the optical and HI velocities are in good agreement.  

\begin{table}
\caption
{A comparision of line of sight velocities obtained from NED and measured HI velocities.}
\begin{tabular}{|l|c|c|}
Name  &  NED velocity   &   HI velocity \\
      &  $km$ $s^{-1}$  &   $km$ $s^{-1}$ \\
\hline

F115-001  & 1131  &  1305 \\
F303-023  & 4485  &  4412 \\
F304-013  & 2250  &  2098 \\
F353-003  & 3739  &  3922 \\
F362-027  & 1344  &  1332 \\
F410-001  & 1545  &  1558 \\
F418-059  & 1673  &  1758 \\
F481-018  & 2087  &  2079 \\
F483-019  & 4128  &  4017 \\
F548-020  & 1961  &  1958 \\
\hline
\end{tabular}
\end{table}

The above procedure resulted in a reduction to 84 detections, but it was clear from inspection of the images from the DSS that for the most distant objects (greater than about 6000 $km/s$) confusion was still a problem. The large beam size covered many faint objects not listed in NED, but distinctly possible sources of the HI emission. In fact the HI emission could arise from the combination of a number of sources in the same group. A simulation (see below) of the expected number of sources in a set of random beams indicated that contamination of the sample was possible at a level better than about 1 in 4 for a sample limited to 5000 $km/s$, but that this drops quickly to about 1 in 2 or worse beyond 6000 $km/s$. Given that our data is not from a set of random sight lines we should expect to do better than this and so (rather arbitrarily) we set a maximum velocity limit of 5500 $km/s$. This fits in well with the previously measured galaxies of table 1, which all have confirmed redshifts below this limit. 

\begin{figure}
\centerline{\psfig{figure=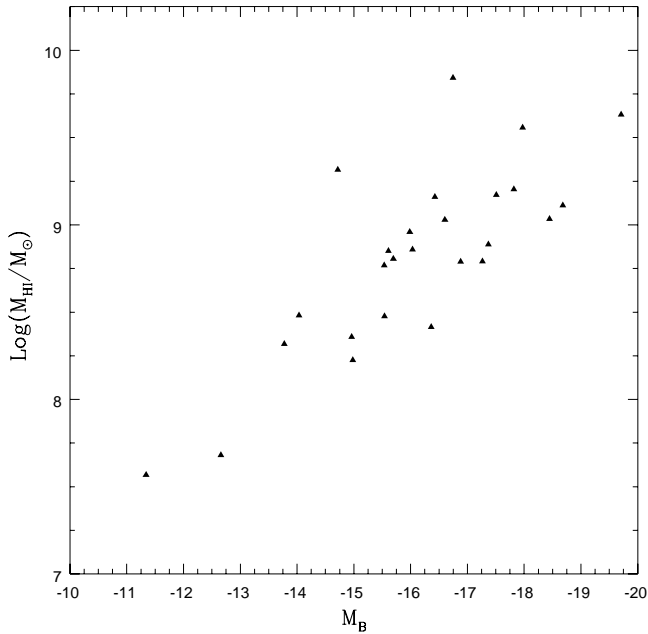,width=9cm}}
\caption
{The HI mass against the absolute B band magnitude}
\label{fig:mvmhi}
\end{figure}

This final sample consists of just 26 objects, out of an initial sample of 2400, that have both a reasonably secure optical and HI detection. Given the above discussion and that our sample consists of relatively isolated galaxies (see below) we believe that our detections are secure and not due to other nearby objects. In  figure~\ref{fig:mvmhi} we have plotted the HI mass against the absolute B band magnitude. As one might have hoped there is a  correspondence between the two, supporting our contention that the optical and HI detections belong to the same objects.

At our sample limits we would expect to be able to detect $\approx 2\times10^{7}$ $M_{\odot}$ of hydrogen at our minimum velocity of 400 $km$ $s^{-1}$ and $\approx 3\times10^{9}$ $M_{\odot}$ of hydrogen at our maximum velocity of 5500 $km$ $s^{-1}$. Given that a 'typical' $M^{*}$ galaxy (like the Milky Way) has $M_{HI}\approx10^{10}$ $M_{\odot}$ 
\cite{zwa97}
 we can detect galaxies that are gas poor compared to $M^{*}$ over our full range of velocities.

Although the low number of combined optical and HI detections is disappointing, it is what we might have expected if our original, optical, selection was sound. The optical selection was designed to select gas poor dE galaxies and this is what it appears to have predominately done. The only other explanation for the low number of HI detections would be that most of the undetected galaxies are at large distances ($v>12000$ $km$ $s^{-1}$), but this is unlikely given the way the optically detected galaxies appear to cluster around nearby brighter galaxies \cite{mor99a}.

\section[]{The spatial distribution of the gas rich galaxies}

\begin{figure}
\centerline{\psfig{figure=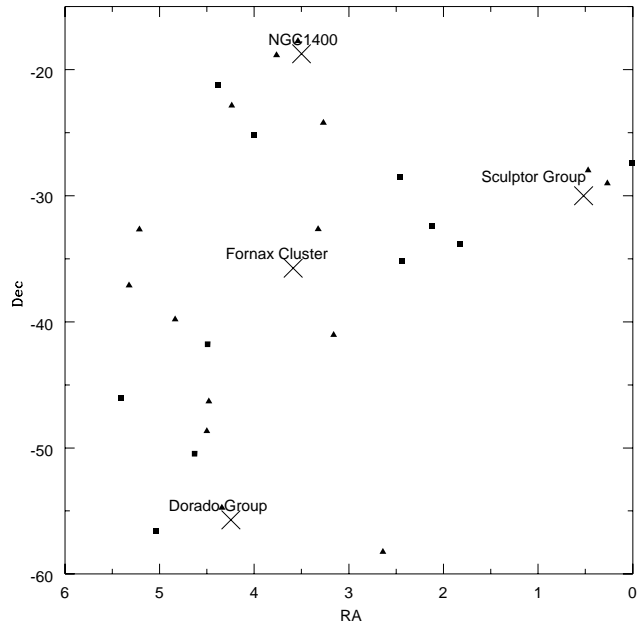,width=9cm}}
\caption
{The position of the HI rich galaxies in relation to nearby groups and clusters. Triangles mark galaxies with velocities less than 2250 $km$ $s^{-1}$, squares those with velocities greater than 2250 $km$ $s^{-1}$}.
\label{fig:pos}
\end{figure}

In figure~\ref{fig:pos} we have plotted the positions of the detected objects compared to the positions of the four major groups/clusters in the survey area. The spatial distribution of our detections is very different to that of the complete optical sample (see fig. 8 in Morshidi-Esslinger et al. 1999a). The optical galaxies cluster about the group/cluster centres while the HI detections appear to avoid the cluster centres almost completely. This segregation of gas rich galaxies from the gas poor has of course been known for some time. It is in low galactic density environments that we would expect to find galaxies like this \cite{sol00}. It is possible that the effect has been enhanced in this case by increased 'optical confusion' due to the higher density of galaxies in the group/cluster centres.

\begin{figure}
\centerline{\psfig{figure=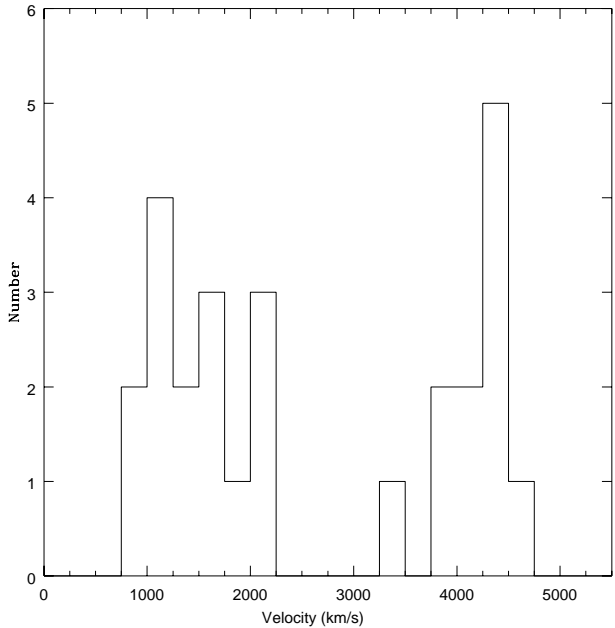,width=9cm}}
\caption
{The distribution of galaxy velocities.}
\label{fig:vel}
\end{figure}

In (figure~\ref{fig:vel}) we have plotted a histogram of the line of sight velocities. The detections cluster at about the velocities expected for
the bright galaxies. Fornax, Dorado, NGC1400 and Sculptor groups/clusters all have redshifts below 2000 $km$ $s^{-1}$. Both Jones and Jones (1980) and Fairall (1998) show that over this region of sky there is a peak in the number density of galaxies at about 1500 $km$ $s^{-1}$ and then a void out to about 4000 $km$ $s^{-1}$. This shows that although these galaxies avoid group/cluster centres they are still associated with the larger scale structure defined by the brighter galaxies. 

\section[]{Mass to Light ratios}

HI masses can be derived using 
\begin{equation}
M_{HI}=2.4 \times 10^{5}d^{2} \int S_{v} dv
\end{equation}

where $M_{HI}$ is the mass of HI in solar units, $d$ is the distance to the galaxy in $Mpc$, $S_{v}$ is the flux density and the integral is over velocity. The flux integral is solved using the Gaussian parameters of the best fitting template as described above.
Distances are obtained by converting line of sight velocities to velocities relative to the Local Group \cite{yah77}. 

We have calculated stellar masses from the absolute blue magnitude of the galaxy assuming a stellar mass to light ratio of one (as suggested by Stavely-Smith et al. (1990) and de Blok et al. (1996) for gas rich galaxies) and an absolute blue magnitude for the Sun of $M_{\odot}^{B}=5.4$ \cite{ban99}. We found that all of the photographic magnitudes listed in \cite{mor99a} were significantly fainter than the available CCD magnitudes listed in NED. We have used the NED CCD photometry wherever possible and made the photographic magnitudes brighter by the mean of the CCD correction where this was not possible. The mean correction was 0.2 magnitudes.

\begin{figure}
\centerline{\psfig{figure=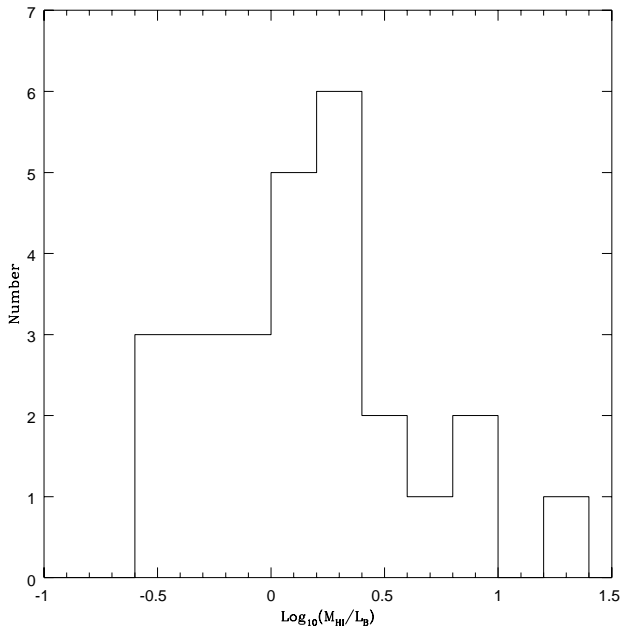,width=9cm}}
\caption
{The distribution of $M_{HI}/L_{B}$ for the HI/optical sample}.
\label{fig:histml}
\end{figure}

In figure~\ref{fig:histml} we show the distribution of $(M_{HI}/L_{B})_{\odot}$ for our sample. It is quite clear that this sample consists  of galaxies with extraordinary values of $(M_{HI}/L_{B})_{\odot}$ (all of the 8 galaxies with $(M_{HI}/L_{B})_{\odot}<1$ have $(M_{HI}/L_{B})_{\odot}>0.3$). The relative gas mass of these galaxies is far higher than that of a 'typical' spiral galaxy \cite{kna90}.

\begin{figure}
\centerline{\psfig{figure=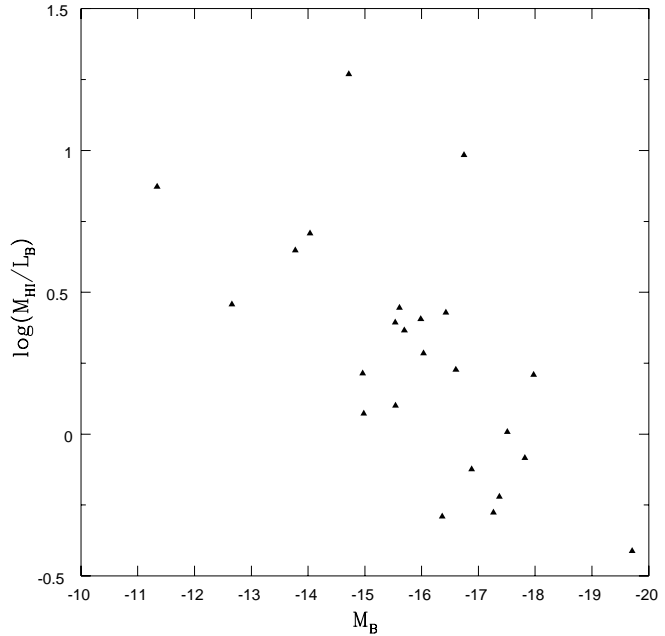,width=9cm}}
\caption
{$M_{HI}/L_{B}$ against absolute blue magnitude for the HI/optical sample}.
\label{fig:mvml}
\end{figure}

In figure~\ref{fig:mvml} we show $(M_{HI}/L_{B})_{\odot}$ plotted against $M_{B}$. Although there is some scatter a clear trend exists for the fainter galaxies to have larger values of $(M_{HI}/L_{B})_{\odot}$. A least squares fit gives $(M_{HI}/L_{B})_{\odot} \propto L_{B}^{-0.4+/-0.1}$ an exponent very close to the value of -0.3+/-0.1 found by Stavely-Smith et al. (1992) for a sample of HI rich dwarf galaxies, but the \cite{sta92} sample was optically selected and has much lower values of $(M_{HI}/L_{B})_{\odot}$ than this sample.

In the same way that selection at any wavelength predominantly selects objects that are bright at that wavelength the combination of relatively faint optical sources with HI selection has led to a sample with large values of $(M_{HI}/L_{B})_{\odot}$. Thus, although small, we have constructed a sample of galaxies that apparently have turned only a small fraction of their gas into stars. 

\section[]{Dynamical masses}
We can make an estimate of the dark matter content of these galaxies by comparing the calculated dynamical masses with the total HI and stellar mass. The main problem with this calculation is knowing the 'dynamical state' of the system. That is whether the system is pressure or rotationally supported. Some of the objects appear to be flattened, some are round, some have a double-horned velocity profiles, but most have what appears to be a Gaussian shape. For these reasons we have used the 'indicative dynamical mass' estimator of Roberts (1978) to approximate the dynamical masses. We have measured optical sizes and inclinations (from the semi-minor to semi-major axis ratio) from the DSS images and then used
\begin{equation}
M_{HI}^{Dyn}=15.0 \theta v \left( \frac{\Delta v}{\sin(i)}\right)^{2}
\end{equation}
where $\theta$ ($arc$ $sec$) is the angular size, $v$ ($km/s$) the line of sight velocity, $\Delta v$ ($km/s$) the velocity width and $i$ the inclination (equation adapted from Banks et al. 1999).

\begin{figure}
\centerline{\psfig{figure=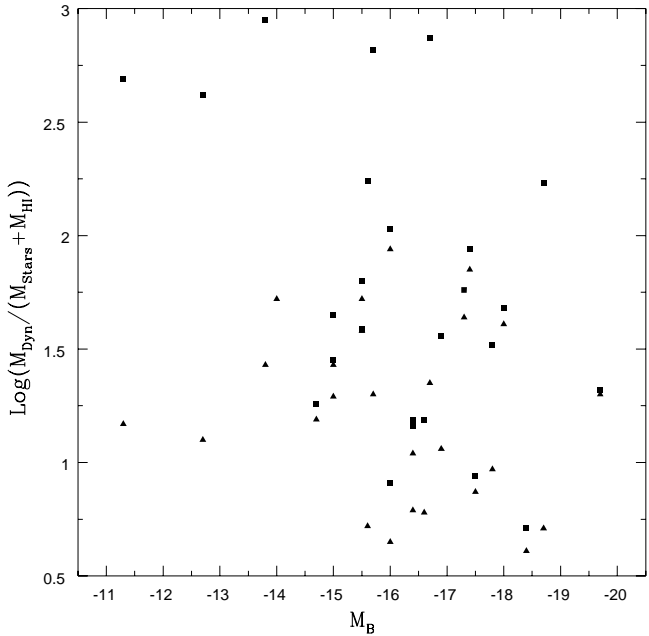,width=9cm}}
\caption
{The indicative dynamical mass to total mass (HI+stars) against the B band absolute magnitude. Squares indicate inclination corrected velocities, triangles no inclination correction (pressure supported).}
\label{fig:dynmasses}
\end{figure}

In figure~\ref{fig:dynmasses} we show the dynamical mass ratio as a function of absolute magnitude. The squares are the calculated values taking into account the $sin(i)$ factor for the velocity widths (a rotating system) while the triangles are with $sin(i)=1$, this is essentially the pressure supported case. For some cases the $sin(i)$ correction is not large and typically the dynamical masses are of order 10 times the mass in gas and stars. This is about the same value that more typical, lower relative gas mass, galaxies have. For a few galaxies the $sin(i)$ correction is large and probably not appropriate. For example the two faintest objects appear as small spheroidal systems and they have Gaussian velocity profiles. The $sin(i)$ correction changes their measured dynamical masses by about a factor of 30. We conclude that isolated galaxies which have only converted a small fraction of their gas mass into stars are dominated by dark matter in a similar way to other galaxies.

\section[]{HI column densities}
So why have these galaxies only converted a small fraction of their HI gas into stars ? Given their apparent rather isolated positions compared to the optically selected sample an obvious solution is that it is interactions with other galaxies that promotes star formation and rapidly consumes gas. Galaxies can enhance their star formation rates if encounters compress the gas in a similar and additional way to the spiral density wave which promotes star formation in spiral galaxies. Our HI sample galaxies appear to be rather isolated and so they must rely on their own internal dynamics to produce the instabilities in the gas that lead to star formation.

\begin{figure}
\centerline{\psfig{figure=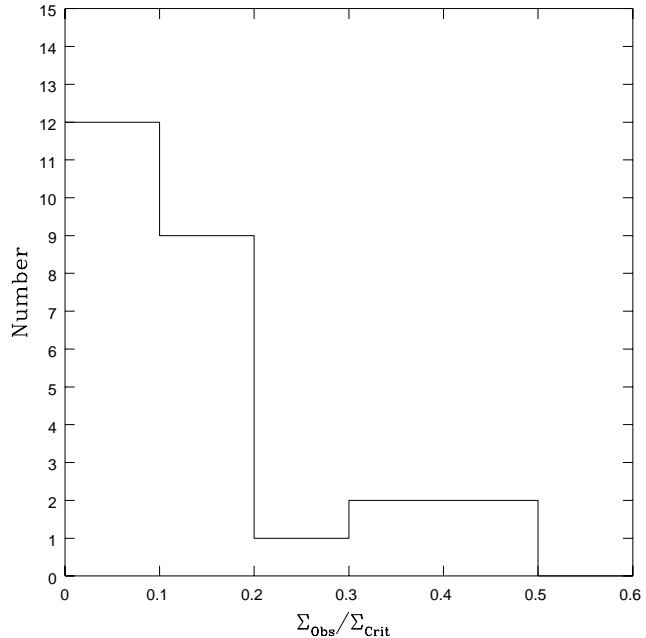,width=9cm}}
\caption
{The ratio of observed HI column density ($\Sigma_{Obs}$) to critical HI column density ($\Sigma_{Crit}$).}
\label{fig:colden}
\end{figure}

Toomre (1964) has proposed an instability criterion for uniformly rotating gas disks (Note that not all of the sample galaxies may be discs). This is a critical column density below which star formation is surpressed. We can calculate the Toomre critical column density ($\Sigma_{Crit}=\frac{3 \times 10^{23} \Delta v^{2}}{\theta v}$, adapted from Kilborn et al. 2000) and compare it with the observed mean HI column density of our sample galaxies (figure~\ref{fig:colden}) assuming that they are uniformally rotating discs. To derive the mean column density we have used the calculated HI mass and the measured optical size. In all cases the observed HI column density is far less than the calculated critical density. Thus these galaxies seem to be deprived of both of the prime ingredients of star formation. They have relatively low mean HI column densities and they are isolated from their companions. 

Given the huge spaces between the galaxies one can speculate on the numbers of HI clouds that suffer a similar, but more extreme, fate than the objects in our sample. 

\section[]{The HI mass function}

There are far too few galaxies in the optical/HI sample to derive an HI mass function in the normal way, but we can infer the relative numbers of low and high HI mass galaxies in another way. One of the great advantages of HI observations, compared to imaging, is that we not only obtain HI masses but also velocities (distances) for the objects. The distribution of distances of detected objects can be used to make an estimate of the slope of the HI mass function. If there are large numbers of massive galaxies compared to low mass then there will be a relatively large number of detections at large distances. Alternately if most of the systems are of low mass then there will be a relatively large number of detections nearby. There are a number of advantages to this method. Firstly we do not have to measure HI masses, we only require detections. We can use our automated method to detect objects above our well defined survey limits - we do not have to rely on detection by eye and a subsequent mass measurement. There are no difficult volume corrections to be made to each mass interval. We can use just the relative numbers at each distance to infer the relative importance of high and and low HI mass galaxies. On the down side, if we want to derive a more qualative value for the low mass slope of the mass function we will still have to compare the observations with a rather idealised model. 

In this section we describe an initial, rather simple, approach to doing this both with regard to interpreting the observational data and modelling the results. The extensive HIPASS data warrents a much more detailed analysis of its noise properties. We will assume that the noise is gaussian. To a good approximation this is true (de Blok, in preparation), but even small deviations from gaussian noise may alter considerably our conclusions. A simple model for comparision with the observations is described below.

The volume $V$ (in $Mpc^{3}$) that a mass of hydrogen $M_{HI}$ can be detected in is
\begin{equation}
V=\frac{1}{12} \pi \theta^{2} \left( 
\frac{M_{HI}}{2.4\times10^{5} \int S_{v} dv} \right) ^{3/2}
\end{equation}
Where $\theta$ ($radians$) is the angular diameter of the beam  and the flux integral corresponds to our survey limits. If the HI mass function has a power law form
\begin{equation}
N(M_{HI}) dM_{HI}=\phi \left( \frac{M_{HI}}{M_{HI}^{*}} \right)^{\alpha} dM_{HI}
\end{equation}
then the total number of detections to some upper mass limit $M_{HI}^{max}$ is
\begin{eqnarray}
N(M_{HI})_{Tot} dM_{HI}\propto &  \int_{0}^{M^{min}_{HI}} M^{\alpha+3/2} dM \nonumber \\
                 & + M^{min^{3/2}}_{HI}
                   \int_{M_{HI}^{min}}^{M_{HI}^{max}} M^{\alpha} dM 
\end{eqnarray}
$M_{HI}^{min}$ is the minimum detectable HI mass at a distance $d_{m}$ and it can be calculated using equation 7. The first integral gives the total number of galaxies with $M_{HI}<M_{HI}^{min}$ detected within a distance $d_{m}$, while the second gives the number of galaxies with $M_{HI}>M_{HI}^{min}$ detected within a distance $d_{m}$.
Assuming that $M_{HI}^{min} << M_{HI}^{max}$ and that $-2.5<\alpha<-1$. 
\begin{equation}
N(M_{HI})_{Tot} \propto (M_{HI}^{min})^{\alpha+5/2}
\end{equation}
so that a plot of the log of the total numbers out to a given distance (velocity) against the log of the distance (velocity) will have slope $2\alpha+5$ because $M_{HI}^{min} \propto d^{2}_{m}$. For example if $\alpha=-2$ then it should have a slope of 1.
If we assume that our initial 2435 sight lines are random then we can use them to carry out the above test by comparing the data with our model. Thus we can use all 155 detections rather than just those with confirmed optical identifications. 

\begin{figure}
\centerline{\psfig{figure=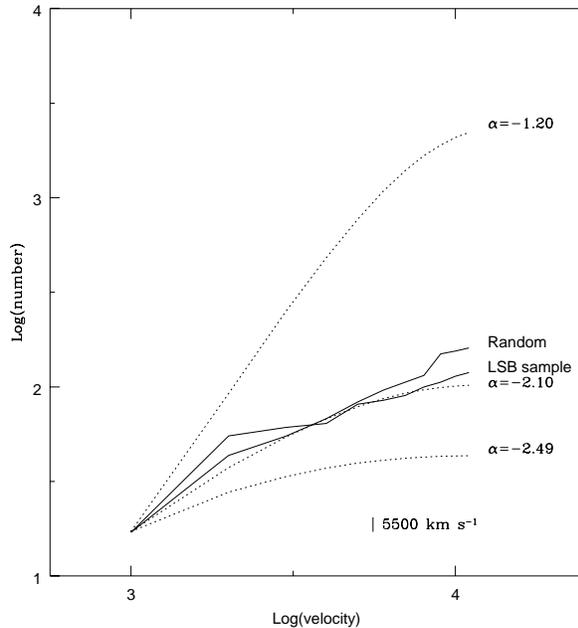,width=9cm}}
\caption
{The cumalative distribution of numbers detected against velocity. Various models are shown with dashed lines, the data as a solid line. The Zwaan et al. HI mass function has $\alpha=-1.2$.}
\label{fig:veldist}
\end{figure}

We have done this after slightly modifying the model mass function form to that of the well known Schechter function and then solving the above integrals numerically. We have used $M^{*}=10^{10}$ $M_{\odot}$ \cite{zwa97} in all of the simulations, changing only the value of the faint-end slope $\alpha$. The result of comparing the distance distribution of our initial 155 detections with various models, each with different faint-end slopes, is shown in figure~\ref{fig:veldist}. Each model has been normalised to the value of the data at a velocity of 1000 $km/s$. We have also indicated on figure~\ref{fig:veldist} a velocity of 5500 $km$ $s^{-1}$, the velocity below which we believe we can optically identify galaxies reliably. The different models are still well separate at velocities below this.

For comparision we have plotted the prediction of Zwaan et al. (1997) who derived the HI mass function from a HI survey using the Arecibo telescope. We infer a faint-end slope much steeper than they derive, though the Zwaan et al. value is typical of a number of other recent determinations \cite{ban99,hen00}. In our sample there are relatively too few detections at higher velocities (or alternately too many at low velocities) for a luminosity function this flat. Our value is consistent though with the steep faint-end slope derived by Schneider et al. (1998). Similarly steep values like this, for the mass function (dark matter), are also predicted by models of hierarchical structure formation \cite{fre96,kau97}

\begin{figure}
\centerline{\psfig{figure=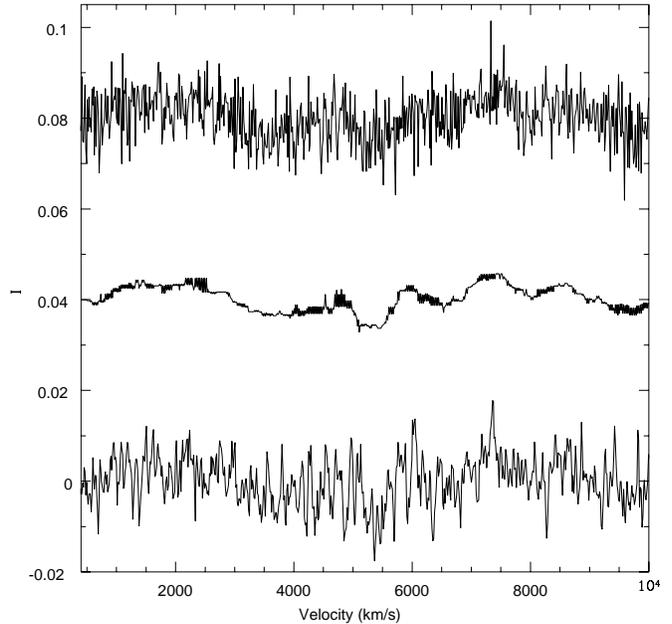,width=9cm}}
\caption
{An example of an HI spectrum after median smoothing and the addition of random noise. The lower spectrum is before smoothing, the middle spectrum is the result of median smoothing and the upper spectrum has had random noise added. The middle and upper specta have been shift in the I direction by 0.04 and 0.08 respectively for clarity.}
\label{fig:noise}
\end{figure}

Our inferred value for $\alpha$ could wrong for a number of reasons
\begin{enumerate}
\item The original optical detection was optimised for nearby galaxies. The simulation assumes a uniform distribution of galaxies while the data were not uniform covering an area of sky containing a number of nearby groups/clusters. If we have selected more nearby galaxies than expected in a random sample then this would mimic a steep mass function faint-end slope in our test. To check this we obtained a further 2500 spectra positioned randomly over the whole southern sky and repeated the cross-correlation analysis. The result was 82 detections. This is 73 less than in the previous sample, confirming our previous view that a large number of the original detections were not associated with the optical galaxies. The random sample predicts a very similar faint end-slope ($\alpha \approx -2$) to that of the optical sample (see figure~\ref{fig:veldist}). Thus the inferred steep faint-end slope does not appear to be due to the original optical selection of the targets. The distribution of velocities for the random sample is quite different to that of the original optical sample. It does not show any noticable clustering features. This is not surprising given the large area covered. The Fairall (1998) maps show galaxies at pretty much all velocities if you consider an area as large as this.
\item The noise is not gaussian so that the the 4$\sigma$ events are noise spikes randomly distributed along the spectrum. This is difficult to rule out but if they were uniformly distributed along the spectrum then the cumlative number detected would just be proportional to $d_{m}$ so we would infer a value of $\alpha=-1.25$. Much flatter than we have measured. If the noise is not uniformly distributed it is more difficult to model, but we can try and assess the probability that these 4$\sigma$ fluctuations occur by chance, given the combination of random and baseline fluctuations. If the noise was purely Gaussian then our minimum detection corresponds approximately to one pixel at 4$\sigma$ and 2 at 2$\sigma$. The probability of this happening by chance is about $10^{-8}$. Given that we have a little over $10^{6}$ pixels in the random data set we would not expect any detections by chance. On the basis of Gaussian statistics 82 detections is a highly significant result. We have tried to quantify the effect the baseline fluctuations may have on this conclusion. By median filtering the spectra (filter width of $\approx500$ $km$ $s^{-1}$) to try to remove any 'real' detections. This should then leave us with just the baseline fluctuations. We have then added Gaussian noise to this with a standard deviation the same as that measured using a pre-filtered baseline subtracted spectrum. This is illustrate in figure~\ref{fig:noise}.  To quantify the uncertainty from baseline noise we repeated the cross-correlation of the random sight-lines using the simulated  smoothed and gaussian noise added spectra. There was only one detection. Thus 82 detections is still highly significant compared to our expectation from the 'model' spectra.
\item At first sight one of the most convincing pieces of evidence that these detections are real is the distribution of velocities of the original 155 detections from the HI/optical sample. We have already shown that the secure optical detections cluster in velocity in the same way as the known large scale structure over this same area of sky (fig. 5). Figure 12 shows that the 4$\sigma$ HI detections from the original HI/optical sample cluster, in velocity, in just the same way, as optically selected galaxies \cite{jon80,fai98}. There are strong peaks at $\approx1500$  and 4500 $km$ $s^{-1}$ and 105 out of 155 HI detections have velocities less than 5500 $km$ $s^{-1}$. In addition the smallest velocity width detections ($\Delta v < 50$ $km$ $s^{-1}$) also cluster in the same way (dotted lines fig. 12), these are the ones most likely to be noise ? The problem is that nearby bright galaxies can 'appear' in beams at some considerable distance away (see the case of NGC1291 and 1365 described in section 3). Given that the region observed contains many nearby groups/clusters this may be a problem though as described in section 3 it is not obvious in any of these cases which galaxy might be the cause of the signal. The cumalative effect of the surface density nearby bright galaxies on the interpretaion of random sight line data requires further investigation. This is something we shall be pursuing further in the future.
\end{enumerate}

\begin{figure}
\centerline{\psfig{figure=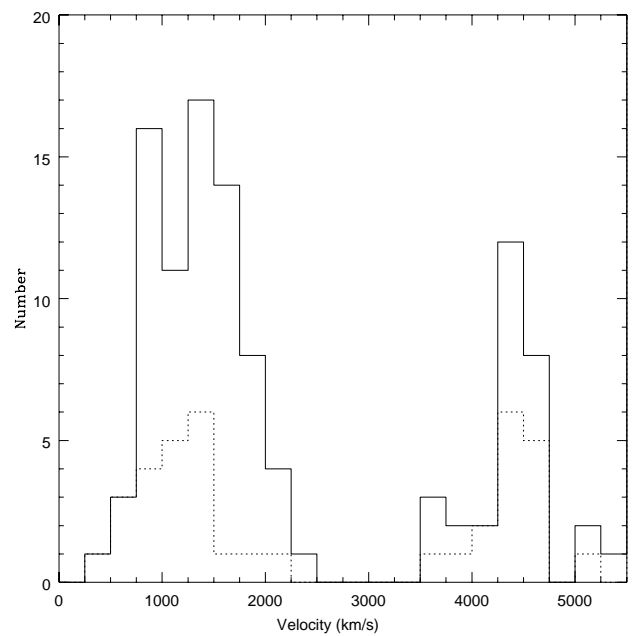,width=9cm}}
\caption
{The distribution HI velocities from the complete HI/optical sample (solid lines). The dotted line is for detections with $\Delta v < 50$ $km$ $s^{-1}$.}
\label{fig:vel2}
\end{figure}

How do these detections compare with other observations at 21cm ? Recently Kilborn (2001) has used a large subset of the HIPASS data to derive the HI mass function. She surveyed $\approx10^{6}$ $Mpc^{3}$ and detected 533 galaxies. Scaling this by the $\approx6\times10^{4}$ $Mpc^{3}$ surveyed by our 2500 random beams we would expect about 32 galaxies in our sample. We actually detected more than twice this number using our automated technique. Looking at this a slightly different way Kilborn also lists the parameters of the derived mass function. Using this data we would predict there to be about 20 $M*$ ($\approx10^{10}$ $M_{\odot}$) galaxies in our random beams. There are actually 22 galaxies with $M_{HI}$ between $5\times10^{9}$ and $5\times10^{10}$ $M_{\odot}$ in the random sample. The two data sets are consistent for $M*$ galaxies.

These results are also not inconsistent with quasar absorption line statistics. According to Rao and Briggs (1993) we should expect $\approx$10 damped $Ly_{\alpha}$ lines in 1000 sight lines to quasars at z=0.65 ($q_{o}=0.5$). If we treat our 2500 sight lines as pencil beams (to give a lower limit) to $\approx12000$ $km$ $s^{-1}$ then we would expect 14 damped $Ly_{\alpha}$ systems in our random sight lines. As the survey column density limit is $\approx10^{19}$ $atoms$ $cm^{-2}$ an order of magnitude less than that required for a damped $Ly_{\alpha}$ system we would also expect to find approximately $14\times(10)^{1.67} \approx 650$ Lyman limit systems \cite{rao93}. Using the numbers against redshift relation of Strengler-Larrea et al (1995), which includes Lyman limit system evolution we come to a similar number ($\approx300$). In fact we have found less than we might have expected from quasar absorption line statistics. This highlights a discrepancy between 21cm and qso absorber observations. Either the column density dependence of the frequency of occurence of quasar absorption lines is strongly evolving or most of the lines fall below our survey limits. Typically Lyman limit absorption lines have measured velocity widths of 10-30 $km$ $s^{-1}$, but of course the line-of-sight only passes through a small part of the object. If the Lyman limit systems are pressure supported then 10-30 $km$ $s^{-1}$ may well be the typical velocity dispersion of the gas and we would only detect those with the largest line widths. If they are rotationally supported then perhaps we are only detecting those that are sufficiently face-on to have high enough central intensities to be detected. In either case these observations are not inconsistent with absorption line studies of quasars. 39 of the 82 detections in the random sample have line widths of 50 $km$ $s^{-1}$ or less.

In the same way as we tried to find corresponding optical and HI detections for the LSB sample we have also tried to identify optical counterparts of the random HI detections. As described before this is very difficult for distant sources, but for nearby objects we might hope that they are both larger and brighter. There is a clear distinction between the optical identification of small
(less than 50 $km$ $s^{-1}$) and large velocity width objects. Nearby large velocity width objects are invariably associated with a bright galaxy, the M* galaxies that we might have expected to detect (see above), small velocity widths with apparently blank fields on the DSS.

\section[]{Conclusions}

Our initial intention was to try and find massive HI galaxies, like Malin 1, by looking at the HI properties of a sample of LSB galaxies. No objects as extreme as Malin 1, with regard to total HI mass, have been found. We have detected a population of extremely gas rich galaxies. These galaxies all have masses within the range of previous well studied galaxies. One striking feature is the very high HI mass compared to stellar mass. These galaxies are either still accumulating gas, they are young and/or they are forming stars at a very slow rate. Detailed (interferometric) observations are required to investigate these alternatives further. The detected galaxies reside in regions of low galactic densities where presumably they have had little interaction with other galaxies. If galaxies had first been detected in HI, rather than the optical, then we would have inferred a very different spatial distribution on the sky, one that was full of voids rather than clusters. Thus HI surveys not only select objects with extraordinary HI properties they also define a very different large scale structure. How extensive this HI rich population is, is  still an open question that warrants a much more detailed analysis of the HIPASS data.


\begin{thebibliography}{99}

\bibitem[Banks et~al. 1999]{ban99}
Banks et al., 1999, ApJ, 524, 612

\bibitem[Barnes et~al. 2001]{bar01}
Barnes et al., 2001, MNRAS, submitted

\bibitem[Bertin \& Arnouts 1996]{ber96}
Bertin E. \& Arnouts S., 1996, A\&AS, 117, 393

\bibitem[Blain et~al. 1999]{bla99}
Blain A., Smail I., Ivison R. \& Kneib J., 1999, MNRAS, 302, 632

\bibitem[Blitz et~al. 1999]{bli99}
Blitz L., Spergal D., Teulen P., Hartman D. \& Burton W., 1999, ApJ, 514

\bibitem[Bothun et~al. 1987]{bot87}
Bothun G., Impey C., Malin D. \& Mould J., 1987, AJ, 94, 23

\bibitem[Bowen et~al. 2000]{bow00}
Bowden D., Tripp T. \& Jenkins E., 2000, Astroph/0011134

\bibitem[Cohen 2000]{coh00}
Cohen J., 2000, Astroph0012109

\bibitem[Davies 1990]{dav90}
Davies J., 1990, MNRAS, 244, 8

\bibitem[Davies et~al. 1994]{dav94}
Davies J., Disney M. \& Phillipps S., 1994, MNRAS, 269, 349

\bibitem[DeBlok et~al. 1996]{deb96}
De Blok E., McGaugh S. \& van der Hulst J., 1996, MNRAS, 238, 18

\bibitem[Disney 1976]{dis76}
Disney M., 1976, Nature, 263, 573

\bibitem[Fairall 1998]{fai98}
Fairall A., 1998, In 'Large Scale Structure in the Univers', Pub. Wiley

\bibitem[Frenk et~al. 1996]{fre96}
Frenk C., Evard A., White S. \& Sammers F., 1996, ApJ, 472, 460

\bibitem[Henning et~al. 2000]{hen00}
Henning et al., 2000, AJ, 119, 2686

\bibitem[Impey et~al. 1988]{imp88}
Impey C., Bothun G. \& Malin D., 1988, ApJ, 330, 634

\bibitem[Impey \& Bothun 1989]{imp89}
Impey C. \& Bothun G., 1989, ApJ, 341, 89

\bibitem[Irwin et~al. 1990]{irw90}
Irwin M., Davies J., Disney M. \& Phillipps S., 1990, MNRAS, 245, 289

\bibitem[Jones \& Jones 1980]{jon80}
Jones J. \& Jones B., 1980, MNRAS, 191, 685

\bibitem[Kambas et~al. 2000]{kam00}
Kambas A., Davies J., Smith R., Bianchi S. \& Haynes J., 2000, AJ, 120, 1316

\bibitem[Knapp 1990]{kna90}
Knapp G., 1990, in 'The Inter-Stellar Medium in galaxies', Pub. Kluwer, Ed. H. Thronson \& J. Shull, p.3

\bibitem[Kauffman et~al. 1997]{kau97}
Kauffman  G., Nusser  A. \& Steinmetz  M., 1997, MNRAS, 286, 795

\bibitem[Kilborn et~al. 2000]{kil00}
Kilborn et al., 2000, AJ, 120, 1342

\bibitem[Kilborn et~al. 2001]{kil01}
Kilborn V., PhD thesis, University of Melborne

\bibitem[Lilley et~al. 1996]{lil96}
Lilley S., LeFevre O., Hammer F. \& Crampton D., 1996, ApJ, 460, L1

\bibitem[Madau et~al. 1996]{mad96}
Madau et al., 1996, MNRAS, 283, 1388

\bibitem[Matthewson \& Ford 1996]{mat96}
Matthewson D. \& Ford V., 1996, ApJs, 107, 97

\bibitem[Morshidi-Esslinger et~al. 1999a]{mor99a}
Morshidi-Esslinger  Z., Davies  J.~I. \& Smith  R.~M., 1999a,
  MNRAS, 304, 297

\bibitem[Morshidi-Esslinger et~al. 1999b]{mor99b}
Morshidi-Esslinger  Z., Davies  J.~I. \& Smith  R.~M., 1999b,
  MNRAS, 304, 311

\bibitem[Phillipps \& Davies 1993]{phi93}
Phillipps S. \& Davies J., 1991, MNRAS, 251, 105

\bibitem[Disney \& Phillipps \& 1983]{dis83}
Disney M. \& Phillipps S., 1983, MNRAS, 205, 1253

\bibitem[Press \& Schechter 1974]{pre74}
Press  W., \& Schechter  P., 1974, ApJ, 187, 425

\bibitem[Rao \& Briggs 1993]{rao93}
Rao S. \& Briggs F., ApJ, 419, 515

\bibitem[Roberts 1978]{rob78}
Roberts M., 1978, AJ, 83, 1026

\bibitem[Schneider et~al. 1998]{sch98}
Schneider S., Spitzak J. \& Rosenberg J., ApJLet, 507, L9

\bibitem[Solanes et~al. 2000]{sol00}
Solanes et al., 2001, ApJ, in press

\bibitem[Stavely-Smith et~al. 1990]{sta90}
Stavely-Smith et al., 1990, AJ, 364, 23

\bibitem[Stavely-Smith et~al. 1992]{sta92}
Stavely-Smith L., Davies R. \& Kinman T., 1992, MNRAS, 258, 334

\bibitem[Stengler-Larrea et~al. 1995]{ste95}
Stengler-Larrea et al., 1995, ApJ, 444, 64

\bibitem[Toomre 1964]{tom64}
Toomre A., 1964, ApJ, 139, 1217

\bibitem[Yahil et ~al. 1977]{yah77}
Yahil A., Tammann G. \& sandage A., 1977, ApJ, 217, 903

\bibitem[Zwaan et~al. 1997]{zwa97}
Zwaan M., Briggs F., Sprayberry D. \& Sorar E., 1997, ApJ, 490, 173

\end{thebibliography}
\end{document}